\documentclass[conference]{IEEEtran}
\IEEEoverridecommandlockouts
\usepackage{cite}
\usepackage{amsmath,amssymb,amsfonts}
\usepackage{algorithmic}
\usepackage{graphicx}
\usepackage{textcomp}
\usepackage{xcolor}
\def\BibTeX{{\rm B\kern-.05em{\sc i\kern-.025em b}\kern-.08em
    T\kern-.1667em\lower.7ex\hbox{E}\kern-.125emX}}

\usepackage{booktabs}
\usepackage{moreverb}
\usepackage{fontenc}
\usepackage{amsmath}
\usepackage{fancybox}
\usepackage{color}
\usepackage{colortbl}
\usepackage{array}

\usepackage{multirow}
\usepackage{multicol}
\usepackage{listings}
\usepackage{graphicx}

\usepackage{subfloat}
\usepackage[captionskip=1pt,farskip=1pt,position=below]{subfig}
\usepackage[export]{adjustbox}
\usepackage{boxedminipage}

\usepackage[ruled,vlined,lined,commentsnumbered]{algorithm2e}
\usepackage{balance}
\usepackage{csvsimple}
\usepackage{longtable}
\usepackage{booktabs}
\usepackage[skip=1pt,labelfont=bf]{caption}
\usepackage{url}
\usepackage{lscape}
\usepackage{rotating}
\usepackage{tikz}
\usepackage[normalem]{ulem}
\usepackage{enumitem}
\usepackage{setspace}
\usepackage{hanging}

\usepackage[most]{tcolorbox}

\usepackage{threeparttable}

\usepackage{marvosym}

\usepackage{enumitem}

\setlist{noitemsep} 

\definecolor{yellow}{RGB}{255,255,153}
\definecolor{grey}{RGB}{224,224,224}

\newboolean{showcomments}
\setboolean{showcomments}{true}
\ifthenelse{\boolean{showcomments}}
 { \newcommand{\mynote}[2]{
      \fbox{\bfseries\sffamily\scriptsize#1}
        {\small$\blacktriangleright$\textsf{\emph{#2}}$\blacktriangleleft$}}}
        { \newcommand{\mynote}[2]{}}

\definecolor{DarkOrange}{rgb}{0.8,0.3,0.0}

\newcommand{\toolname}{{\sc APISens}\xspace}

\begin{document}

\title{\toolname - Sentiment Scoring Tool for APIs with Crowd-Knowledge
}

\makeatletter
\newcommand{\linebreakand}{%
  \end{@IEEEauthorhalign}
  \hfill\mbox{}\par
  \mbox{}\hfill\begin{@IEEEauthorhalign}
}
\makeatother

\author{\IEEEauthorblockN{Kisub Kim}
\IEEEauthorblockA{\textit{Singapore Management University} \\
Singapore \\
kisubkim@smu.edu.sg}
\and
\IEEEauthorblockN{Ferdian Thung}
\IEEEauthorblockA{\textit{Singapore Management University} \\
Singapore \\
ferdianthung@smu.edu.sg}
\and
\IEEEauthorblockN{Ting Zhang}
\IEEEauthorblockA{\textit{Singapore Management University} \\
Singapore \\
tingzhang.2019@phdcs.smu.edu.sg}
\linebreakand
\IEEEauthorblockN{Ivana Clairine Irsan}
\IEEEauthorblockA{\textit{Singapore Management University} \\
Singapore \\
ivanairsan@smu.edu.sg}
\and
\IEEEauthorblockN{Ratnadira Widyasari}
\IEEEauthorblockA{\textit{Singapore Management University} \\
Singapore \\
ratnadiraw.2020@phdcs.smu.edu.sg}
\and
\IEEEauthorblockN{Zhou Yang}
\IEEEauthorblockA{\textit{Singapore Management University} \\
Singapore \\
zyang@smu.edu.sg}
\linebreakand
\IEEEauthorblockN{David Lo}
\IEEEauthorblockA{\textit{Singapore Management University} \\
Singapore \\
davidlo@smu.edu.sg}
}


\maketitle
\begin{abstract}
\label{sec.abstract}
Utilizing pre-existing software artifacts, such as libraries and Application Programming Interfaces (APIs), is crucial for software development efficiency. 
However, the abundance of artifacts that provide similar functionality can lead to confusion among developers, resulting in a challenge for proper selection and implementation. 
Through our preliminary investigation, we found that utilizing the collective knowledge of a crowd can greatly assist developers in acquiring a thorough and complete understanding of the complexities involved in the software development process.
Especially as emotions are an inseparable part of human nature, it influences developers' activities. 
In this regard, we attempt to build a tool that can retrieve sentiment information for software APIs so that developers can determine APIs to utilize for their tasks. 
We employ the dataset from the most popular platforms (i.e., Twitter and YouTube) to build our research prototype. 
The source code, tool, and demo video are available on GitHub at \url{https://github.com/FalconLK/APISens}.
\end{abstract}



\section{Introduction}
\label{sec.introduction}

Over the past decades, we have encountered the rapid growth of open-source software (OSS). 
This phenomenon naturally drives more and more reuse of software artifacts (e.g., libraries or frameworks).
As the community has tremendous demand, artifact recommendation systems have received extensive attention.
For example, developers often search for existing source code artifacts~\cite{sim_how_2011,gallardo_valencia_internet_scale_2009} to obtain the functions they need to implement or to maintain the software~\cite{xia2017developers}.
Therefore, many automated artifact recommendation systems have been proposed~\cite{thung2013automated,wei2022clear,gu2016deep,huang2018api,liu2018effective,nguyen2016api} to help developers select artifacts efficiently.

The reuse of devised software artifacts causes another problem due to there being a large number of artifacts that implement the same or similar functionalities, which may confuse developers. 
For example, two APIs, {\tt getData}
and {\tt retrieveInformation}, are capable of processing the same functionality, which is ``requesting and obtaining the necessary data''. 
As a more complex example, there exists  {\tt decrypt} and {\tt unscramble} whose names are totally different while they are both for ``decoding/cracking something to get the initial state''.
Among the multiple options, developers need to choose the most appropriate ones. 
A study~\cite{xu2020reinventing} suggests that software reuse mainly occurs due to a lack of base knowledge of the corresponding artifacts.
This indicates that developers reuse artifacts because they do not know what they need to know to select the most suitable one.
They also revealed that re-implementation happens when the artifacts are required to be further enhanced, the dependencies are too complicated, or they are deprecated.
Moreover, many developers are actively discussing choosing a suitable artifact in the software development community\footnote{https://stackoverflow.com/questions/488348/what-are-your-criteria-for-choosing-a-framework-or-library} and their websites\footnote{https://www.lagarsoft.com/blog/8-tips-for-choosing-the-right-library}\footnote{https://www.theserverside.com/tip/7-tips-to-choose-the-right-Java-library}.

To support developers with a comprehensive understanding when they select a software artifact, crowd-knowledge is known to be helpful~\cite{biswas2020achieving,zhang2020sentiment}. 
Especially as emotions are an inseparable part of human nature, it influences developers' activities as well~\cite{de2013understanding}.
Several studies in the software engineering community discovered that developers express sentiments on libraries~\cite{xu2020reinventing}, APIs~\cite{uddin2019understanding}, commit messages~\cite{guzman2014sentiment}, project artifacts~\cite{murgia2014developers}, etc.
Yet, most of the existing recommendation systems~\cite{gu2016deep,gu2018deep,thung2013automated,kim2018facoy,wei2022clear} retrieve multiple artifacts without considering their quality in terms of developer sentiments while only 
some approaches~\cite{wei2022clear,cai2019biker} considered discussion data from Stack Overflow\footnote{\url{https://stackoverflow.com/}}.

These observations motivated us to build a tool named \toolname, a software API sentiment scoring tool with crowd-knowledge (i.e., online developer sentiments) from diverse resources.
\toolname retrieves the sentiment scores for APIs such that the user can recognize its popularity, level of awareness, or how the public reacts to it.






To construct the tool, we employ a pre-trained Transformer (i.e., a deep neural network architecture based on the attention mechanism) model, BERT (Bidirectional Encoder Representations from Transformers)~\cite{devlin2018bert}.
Its effectiveness in sentiment analysis has been proved by an empirical study~\cite{zhang2020sentiment}.
\toolname consists of two models that are based on BERT-base. 
The first model is a recognition model, which distinguishes whether the input discussion is related to a software artifact context instead of a normal context (e.g., {\tt decrypt} and {\tt unscramble} may be used for a normal context anywhere).
We collect broad discussion text data from diverse online resources (i.e., Twitter and YouTube) and filter with the first model.
Please note that the discussion with the normal context is the noise for training the API sentiment analysis model. 
Furthermore, we believe that the recognition model, which can distinguish discussions related to software artifacts, is able to extract API-related ones when it meets the discussion that contains API-related content. We call the results of the first model API-related discussions.
The second model infers a score for the input API. 
It is fine-tuned with API-related discussions to retrieve the score for an API by analyzing its sentiments. 
In a nutshell, \toolname takes an API name as text and retrieves a comprehensive sentiment normalized score for such an API.


In summary, this paper contributes the following:
\begin{itemize}
    \item Diversity of the resources (i.e., Twitter and YouTube) for fine-tuning the BERT model to infer the sentiment score that supports users with comprehensive API scores.
    \item Construction of a tool that can provide multiple scores including sentiment and popularity for software APIs such that developers can benefit from determining more appropriate APIs for their tasks.
\end{itemize}

\section{Related Work}
\label{sec.relatedwork}

\subsection{Sentiment Analysis for Software Engineering}
Sentiment analysis in software engineering is a computational study of various viewpoints of developers on diverse software artifacts. 
Most of them consider sentiment analysis as a polarity classification. 
Given a text, the goal of the study is to predict its sentiment orientation among \textit{positive}, \textit{neutral}, or \textit{negative}. 
We introduce representative approaches to the topic.

Stanford CoreNLP~\cite{socher2013recursive} is designed for single-sentence classification; it is trained with the Recursive Neural Tensor Network~\cite{bakshi2016opinion} on the Stanford Sentiment Treebank~\cite{socher2013recursive}. 
SentiStrength~\cite{thelwall2010sentiment} is a lexicon-based approach with several dictionaries, including both formal and informal terms, while each term is labeled with a sentiment strength. 
SentiStrength-SE~\cite{islam2017leveraging} contains a domain-specific dictionary that is constructed based on an in-depth investigation of the results from the previous approach. 
SentiCR~\cite{ahmed2017senticr} is designed for code review comments.
The results show that Gradient Boosting Tree~\cite{pennacchiotti2011democrats} is the most suitable model for their data. 
Senti4SD~\cite{calefato2018sentiment} is a supervised model that utilizes three different features based on (1) generic sentiment lexicons~\cite{thelwall2010sentiment}; (2) keywords (number of occurrences); (3) word representation in a Distributional Semantic Model (DSM) specifically trained on Stack Overflow data.
Recently, a study by Zhang et al.~\cite{zhang2020sentiment} was conducted to assess the performance of these sentiment analysis tools and large pre-trained models that are frequently leveraged within the software engineering domain.

\subsection{Deep learning based API Recommendation}
Recent API recommendation approaches utilize neural networks to learn and identify patterns in API usage and documentation.
These systems can be trained on datasets consisting of past API usage and documentation, allowing them to make informed recommendations for future API selection.

DeepAPI~\cite{gu2016deep} is the first introduced deep learning model to API recommendation, achieving end-to-end API sequence generation.
This idea considered the API recommendation as a machine translation problem with a Recurrent Neural Network (RNN) EncoderDecoder model to encode a query into a context vector. Then, it recommends an API sequence based on the context vector for the query. 
API2Vec~\cite{nguyen2017exploring} utilizes unsupervised learning and word embedding techniques to create embeddings and recommend APIs based on their semantic similarity to a given context.
Huang et al.~\cite{cai2019biker} unveiled BIKER by filtering prospective APIs according to their alignment with Stack Overflow posts by leveraging the bag-of-word embedding to optimize the selection of APIs. 
CLEAR~\cite{wei2022clear} also leverages Stack Overflow posts and BERT sentence embedding to preserve the semantic information in queries and posts.
Moreover, it employs contrastive learning~\cite{chen2020simple} to distinguish the queries which are semantically dissimilar but lexically similar.
Although BIKER and CLEAR took into account the crowd-knowledge in a way with Stack Overflow posts, they only adopted related terms with the target APIs.
Therefore, they still missed considering developer sentiments.

\section{\toolname in Detail}
\label{sec.libsense}
\toolname takes an API name and retrieves two types of sentiment scores; one from discussion and the other from videos. 
As Figure~\ref{fig:overview} illustrates, 
the tool consists of the following three components: (1) API Recognizer, (2) Discussion Analyzer, and (3) Video Analyzer.
Assuming the data from online platforms are crawled with the corresponding APIs (e.g., GitHub REST API\footnote{\url{https://docs.github.com/en/rest?apiVersion=2022-11-28}}), \toolname must take a simple form of API name as a query.
Given an API name as a query, \toolname's pre-trained API Recognizer identifies the API-related discussions (i.e., tweets in this tool).
Once the relevant discussions are detected, Discussion Analyzer takes them as input to retrieve the sentiment scores.
Video Analyzer concurrently analyzes sentiments from the videos corresponding to the target API with various aspects (i.e., the number of `Likes', `Comments', and `Views').
Finally, the user interface shows the results. 
Figure~\ref{fig:overview} illustrates the steps that are unfolded in the working of the score retrieval.

\begin{figure}[!tp]
\begin{center}
\includegraphics[width=0.9\linewidth]{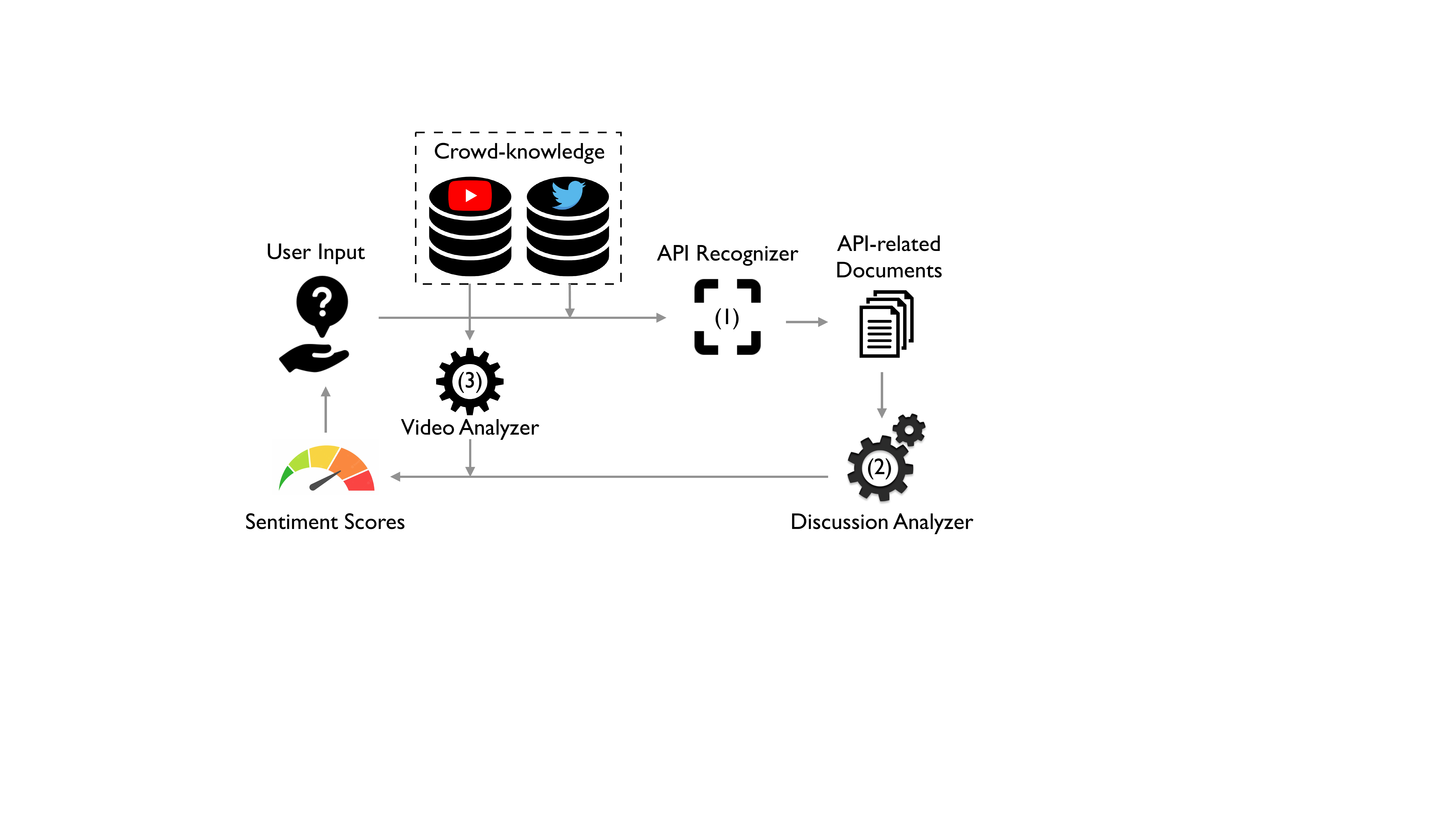}
\caption{Overview of \toolname.}
\label{fig:overview}
\end{center}
\end{figure}

\subsection{Datasets (Crowd-knowledge)}
We collect datasets from two big platforms (i.e., Twitter and YouTube).
Twitter is one of the biggest social network platforms, and a sentiment analysis tool can benefit in many aspects. 
It has a large user base, including developers, which can provide a broad overview of sentiment about software APIs, while its large and diverse discussions can ensure that the tool is able to handle a wide range of sentiment. 
The data from Twitter is constantly being updated, which is useful for practitioners to be up-to-date.
YouTube is the largest video-sharing platform with a wide range of content across many different topics, making it a valuable data source for sentiment analysis.
Many people rely on YouTube to get information and opinions about products, services, and even software APIs.

In order to obtain API datasets from both Twitter and YouTube, we concurrently collect data from the platforms. As tweets may be written in various languages and may contain various forms of noise, such as retweets, mentions of other users, URLs, and emojis, the data is preprocessed prior to collection.
The preprocessing step converts emojis to their corresponding textual representation and filters out APIs and corresponding tweets by calculating the average number of tweets per API (i.e., APIs that are related to less than the average are omitted). This filtering helps to prevent partial generalization, as the inclusion of tweets from APIs with only a single tweet could lead to bias.
Then, it removes duplicate tweets and those that are written in non-English using a Python library (i.e., langdetect). The detection is possible by taking the language with the highest probability.
The video dataset contains the title, description, the number of `Likes', `Comments', and `Views'.
Some videos may not contain the statistics of the information for sentiment analysis, and they are the further candidates to be eliminated.
We initially target 1409 JDK APIs for our research prototype and collected 56,646 tweets. 
After processing the data, the number of tweets is 28,278, with the corresponding 476 APIs.
While the crawler can be performed in real-time and \toolname can analyze the dynamic data, our research prototype assumes the dataset is already collected.

\subsection{API Recognizer}
Our API recognizer is composed of a fully connected layer on top of a pre-trained BERT~\cite{devlin2018bert} as it has shown to be effective in the task of software library recognition in tweets~\cite{9796362}.
We prepend the [CLS] token to each tweet as the input to BERT.
The final hidden state of the [CLS] is seen as the aggregated representation of the tweet.
We use the embedding output from [CLS] as the input to the classifier.
Our API recognizer is trained with the AdamW optimizer~\cite{loshchilov2017decoupled}
 and we use a linear learning rate scheduler.

The employed model is to predict whether the tweets are related to software artifacts or not.
It was initially trained with a library dataset (i.e., 4,456 tweets with 23 libraries). Still, the key features (i.e., tokens related to the software artifacts) that such a recognition model learns are not different for APIs.
Furthermore, the literature~\cite{9796362} shows that the mixed-setting (i.e., utilizing the mixed dataset across the different libraries) performs the best with 90\% of the F1-score.
Therefore, it is a suitable model to leverage with the same setting to classify the discussions related to our target APIs. 
Based on the predictions, the authors manually confirm the labels to enhance the correctness.

Moreover, we classify the video datasets since there may exist noise, such as the videos related to the normal terms such as {\tt unscramble} can be associated with any topic. 
As we have the titles and descriptions, we apply the same procedure to filter out the noise.
As a result of the video filtering procedure, the number of target APIs decreased from 476 to 235, and the number of associated tweets dropped to 1,606.


\subsection{Discussion Analyzer}
Given the API-related discussion dataset, this component performs sentiment analysis to get a score for retrieval.
We again employ the same pre-trained BERT model that is leveraged in a state-of-the-art sentiment analysis approach~\cite{zhang2020sentiment}.
In the literature, the experimental results show that BERT outperforms other state-of-the-art techniques that specifically target sentiment analysis for software engineering downstream tasks (i.e., BERT shows 89\% of Micro-avg, which is 7 percentage points better than the second-best model~\cite{zhang2020sentiment}).
The API dataset used for training was 4,522 Stack Overflow posts related to software APIs.
As the model retrieves class predictions among \textit{Positive}, \textit{Neutral}, and \textit{Negative}, we convert them to integers 10, 5, and 0, respectively, to calculate the average scores with a maximum of 10.
We finally integrate the results and calculate the average sentiment score for every tweet for the input API to provide a comprehensive and concise score. 

\subsection{Video Analyzer}
Given the list of API-related videos filtered with their title and description by the API Recognizer, 
\toolname extracts the statistics to support the sentiment more comprehensively. 
The statistics include the number of likes, comments, and views of each video that are leveraged to determine the sentiment of the content (i.e., the input API).
These numbers of a video are known to be indicative of public sentiment toward the video.
A high number of likes and views imply that the video is popular and well-created, while a high number of comments can provide insight into the types of discussions and reactions the video elicits.
We believe these numbers can be useful for sentiment analysis as they directly reflect audiences' opinions that may contain emotional orientations. 
To get a more accurate reflection of the central tendency in the result set while avoiding the outlier effects, we decide to provide each median value in the user interface.

\section{Prototype explanation and Evaluation}
\label{sec.usecase}

\subsection{Prototype Explanation}
The user interface of \toolname is designed to be intuitive and easy to use. 
Figure~\ref{fig:ui} illustrates the user interface.
The main window is divided into three main sections: the query panel on the top, the result demonstration panel in the center, and the chart panel on the bottom.
The query panel (\textcircled{1}) takes an API name as the user query. 
Once the \textit{Search} button is clicked, it analyzes the discussion sentiment to retrieve the score to a result demonstration panel (\textcircled{2}). 
At the same time, it provides the sentiment information from the related videos in \textcircled{3}.
As we mentioned, \toolname calculates the median value of the collected videos of the input API.
Furthermore, a bar chart (\textcircled{4}) displays statistics of likes and comments from a certain number of videos related to the corresponding API in the chart panel.
This allows users to recognize the trend of actual numbers of likes and comments on related videos.
Users can re-sample the number of videos to get more drift using the slide bar (\textcircled{5}).
The bar (\textcircled{6}) at the top demonstrates the progress.

\begin{figure}[!tp]
  \centering
  \includegraphics[width=0.85\linewidth,valign=t]{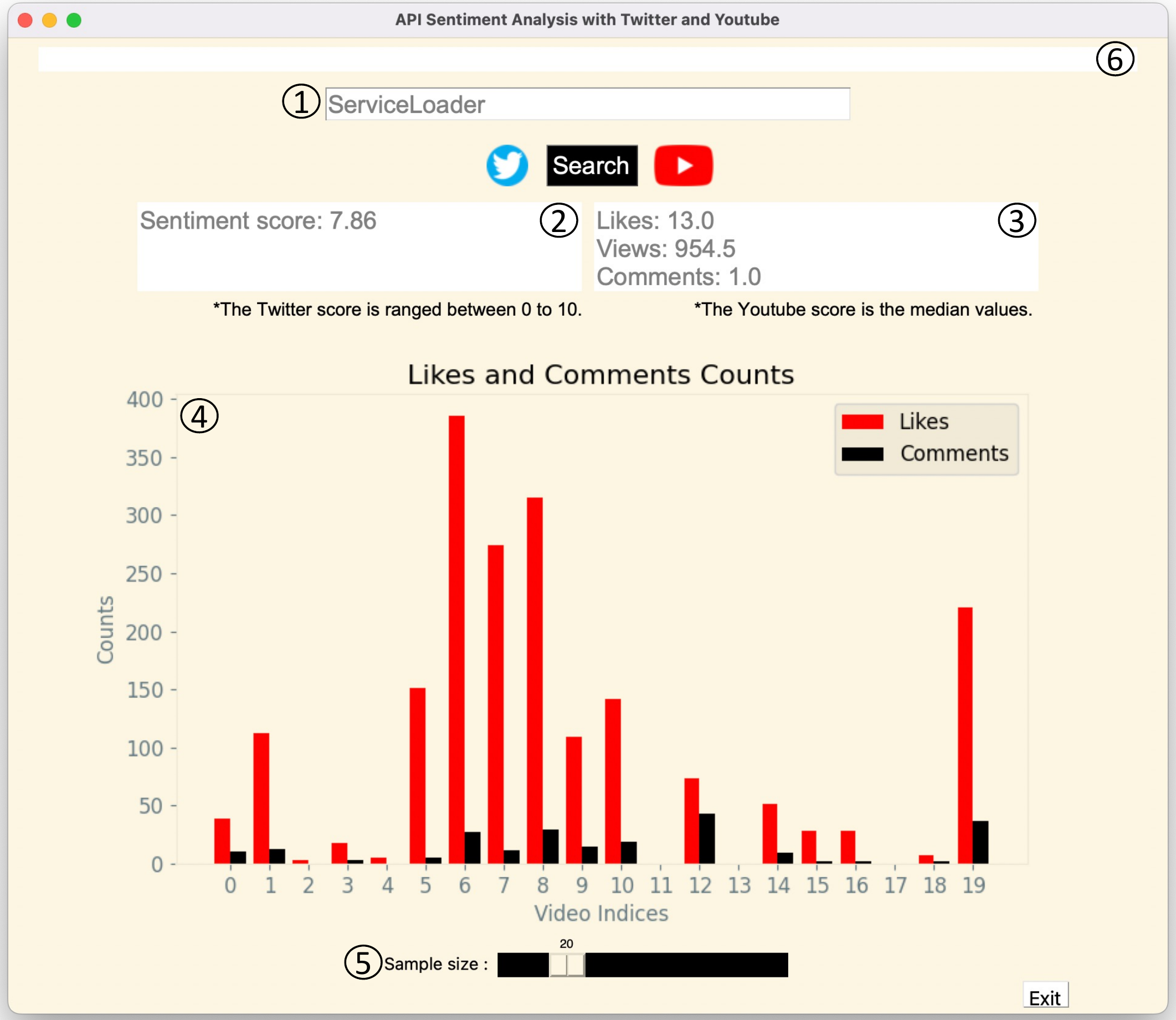}
  \caption{User Interface of \toolname}
  \label{fig:ui}
\end{figure}


\subsection{Possible Use-cases from User Experience}
As our goal is to support practitioners, we also conduct a user study/discussion evaluating \toolname's usefulness.
We ask 2 Research Scientists and 2 Ph.D. students in Software Engineering who have more than 5 years of programming experience.
To obtain opinions, we first deliver \toolname with 20 samples of APIs and let the participants rate (i.e., 1 to 5 where 1 indicates strongly disagree and 5 denotes strongly agree) their usability and usefulness as well as discuss the potential.
Overall, the participants consider \toolname to be easy-to-use (5 out of 5) and useful (4.05 out of 5). 
We have identified and established the promising use cases from the user discussion.


\noindent
\textbf{Understanding the popularity of the API at a shot.}
The main use case for \toolname is, indeed, getting the sentiment score for an API such that the user can recognize whether the API is widely used and well-supported.
Also, it can give a sense of how much demand there is for the API as well as it allows users to get valuable context and perspective when comparing different APIs or considering potential alternatives.

\noindent
\textbf{Incorporating it from an IDE.}
The sentiment information could be used to personalize API recommendations for individual developers based on their preferences and needs. 
Hence, the integration can provide more options.
It also benefits code completion of the IDE, as sentiment information can be a factor for prioritization for real-time API suggestions. 

\noindent
\textbf{Connection with sequence recommendation techniques.}
Once \toolname is connected with API sequence recommendation, it can bring synergies such as prioritize/re-order recommending APIs with high positive sentiment scores.
This could potentially lead to a better user experience.

\noindent
\textbf{Sentiment information as learning features.}
The sentiment information could be used as a feature in machine learning models to improve the accuracy of API recommendations. 
For example, as there is a strong correlation between API sentiment and API popularity or usage, including such information as a feature in a model can potentially improve performance.

\section{Further Ideas}
\label{sec.conclusion}

Based on our observations and early feedback from the users, we believe that \toolname has the potential to be a useful tool for API comprehension.
Here are some ideas to further improvement: \textbf{Concurrent support for multiple APIs.}
Developers tend to compare multiple APIs to find the best API for a particular task or to stay up to date with industry trends. 
This could help them stay informed about new APIs and best practices and ensure that they are using the most popular and effective APIs.
\textbf{Training the models with a bigger dataset.}
To maximize the performance of the tool, the pre-trained models can be fine-tuned with a bigger dataset.
As statistical power and generalizability of the model can be derived by the bigger and better dataset, \toolname can be further fine-tuned for more reliable results.
Moreover, covering more APIs may help our tool to be more generally used by developers. 
\textbf{Further support with similar APIs that perform the same functionality.}
\toolname can encompass a function that can support similar APIs as well as their sentiment information corresponding to the queried one. 
This can further boost efficiency and ease the comparison and determination process for more suitable APIs. 



\balance
\bibliographystyle{IEEEtran}


\end{document}